\documentclass[journal,transmag]{IEEEtran}
\usepackage{cite}
\ifCLASSINFOpdf
\else
\usepackage[dvips]{graphicx}
\fi
\usepackage{amsmath,ragged2e,color,amsmath, amsthm, amssymb,amsfonts}
\usepackage{algorithmic}
\usepackage{array}
\usepackage{subfig}

\begin{document}

\title{Analysis of ion-acoustic rogue wave in complex magneto-plasmas}

\author{\IEEEauthorblockN{A. Parvez\IEEEauthorrefmark{}}
\IEEEauthorblockA{\IEEEauthorrefmark{}Department of Physics, Uttara University, Bangladesh}
\IEEEauthorblockA{\IEEEauthorrefmark{}Email: parvez.asif@uttarauniversity.edu.bd}
\thanks{}}

\IEEEtitleabstractindextext{
\begin{abstract}
  I considered a four-component magnetized plasma medium consisting of opposite polarity ions and
  super-thermal distributed positrons and electrons to investigate the stable/unstable frequency
  regimes of modulated ion-acoustic waves (IAWs) in the D-F regions of Earth's ionosphere and laboratory
  plasmas. A $(3+1)$-dimensional nonlinear Schr\"{o}dinger equation is derived. The parametric regimes for
  the existence of the MI, first- and second-order rogue waves, and also their basic features (viz., amplitude,
  width, and speed) are found to be significantly modified by the effect of physical plasma parameters (such as superthermal
  index and positron to electron temperature ratio) and external magnetic field. It is found that the nonlinearity
  of the different types of electronegative plasma system depends on the positive to negative ion mass ratio. It is also
  shown that the presence of super-thermal distributed electrons and positrons modifies the nature of the MI of the modulated IAWs.
  The implication of our results for the laboratory plasma [e.g., ($Ar^+,~F^-$) electronegative plasma] and space plasma [e.g., ($H^+,~H^-$), ($H^+,~O_2^-$)
  electronegative plasma in D-F regions of Earth's ionosphere] are briefly discussed.
\end{abstract}
\begin{IEEEkeywords}
Complex magneto-plasma, 3D NLSE, stable/unstable regions, rogue wave.
\end{IEEEkeywords}}
\maketitle
\IEEEdisplaynontitleabstractindextext
\IEEEpeerreviewmaketitle
\label{4sec:Introduction}
Electronegative plasma \cite{Vladimirov2003,Misra2008,Tie-Lu2015,Panguetna2018,Sabry2009,Abdelwahed2016,Selim2016,Chowdhury2017}, containing
both positive and negative ions as well as electrons, is one of the fascinating
research topics among the modern plasma physicists because it is not only observed in astrophysical
objects (viz., Earth's ionosphere \cite{Massey1976,Khondaker2019} (D-region \cite{Pedersen1965,Mishra2002,Mishra1995} and F-region \cite{Sabry2009}), Solar wind \cite{Abdelwahed2016}, Titan's
ionosphere \cite{Coates2007}, and cometary comae \cite{Chaizy1991}, etc.) but also in both industrial
use \cite{Panguetna2018} and laboratory experiment \cite{Jacquinot1977}. The antielectron (positron)
may be simultaneously created in many astrophysical environments, such as solar flares \cite{Hansen1988},
interstellar medium \cite{Adriani2009}, and modern laser plasma
experiments \cite{Surko1989,Greaves1994,Chowdhury2017a}.
Sabry \textit{et al.} \cite{Sabry2009} have investigated the propagation
properties of ion-acoustic (IA) solitary waves (IASWs) in a collisionless unmagnetized plasma consisting of
positive-negative ions species as well as electrons. Chowdhury \textit{et al.} \cite{Chowdhury2017} have
examined heavy IA rogue waves (HIARWs) in an unmagnetized plasma medium consisting of positive ions, negative ions, electrons, and positrons and found
that the basic features, such as MI criteria, growth rate, and amplitude of the pulse of HIARWs are
affected by the various plasma parameters.

The linear and nonlinear properties of DAWs totaly depend on the velocity distribution functions of plasma species
which are not thermally equilibrium \cite{Vasyliunas1968,Hellberg2002,Sabry2012}.
Many observational data have confirmed that the presence of fast particles in space,
such as solar wind \cite{Sabry2012}, Earth's bow-shock \cite{Asbridge1968,Cairns1995},
magnetosphere of Jupiter \cite{Leubner1982}, Saturn \cite{Armstrong1983}, in the vicinity of the Moon \cite{Futaana2003},
and auroral zone plasma \cite{Futaana2003}, etc. Such fast particles are described by
the non-Maxwellian Kappa ($\kappa$) distribution which was first introduced by Vasyliunas \cite{Vasyliunas1968}.
In kappa-distribution, the spectral index ($\kappa>3/2$) stands for the
strength of particle super-thermality, and also small value of $\kappa$ stands for strong
deviation from Maxwellian distribution. A number of authors have studied the various
nonlinear waves in plasmas by considering $\kappa$-distributed plasma species \cite{Sabry2012,Chowdhury2017a,Shahmansouri2013,Abdelwahed2016}.
Shahmansouri \textit{et al.} \cite{Shahmansouri2013} have studied the arbitrary amplitude ion-acoustic solitary waves (IASWs) with
two-temperature super-thermal electron and found that the electron super-thermality decreases the phase velocities of both modes.
Later, Abdelwahed \textit{et al.} \cite{Abdelwahed2016} have investigated the propagation of rogue waves in an un-magnetized pair-ion
plasma and found that the amplitude of the rogue waves (RWs) increase with super-thermality index $\kappa$.

Rogue wave is a complex, rare, short-lived and high energetic pulse, which is an interesting research topic for the plasma physicists.
Therefore, researchers have theoretically investigated the properties of rogue waves in
many plasma systems \cite{Abdelwahed2016,Mei2016,Guo2012,Sabry2012}.
Sabry \textit{et al.} \cite{Sabry2012} have reported the propagation properties of IARWs in an
unmagnetized plasma medium with warm ions, electrons and positrons and analyzed that the IARWs
become suddenly high energetic pulse around the critical wave number ($k_c$) and later decrease with
increasing the $k_c$. Abdelwahed \textit{et al.}
\cite{Abdelwahed2016} have studied the RW in collisionless superthermal plasma with opposite polarity
ions as well as superthermal electrons and found that various plasma parameters, such as spectral index
$\kappa$, ions density ratio, and ions mass ratio ($m_+/m_-$), etc., play an important role on the RW
propagation properties.
For better understanding the nonlinear wave phenomena in space and laboratory plasma systems,
it is necessary to take into account the external uniform magnetic field (EUMF) which strongly
affects the propagation properties of the wave. The above-mentioned
studies did not consider the magnetic field which interpenetrates in most of the laboratory and astrophysical plasmas.
Recently, a few authors investigated the properties of RWs in
different plasma systems \cite{Guo2014,Haque2019,Haque2019a,Haque2020,Haque2021,Jahan2021}.

Here, I consider a three dimensional $(3D)$ model in order to also account for the
transverse effects of RWs. I work in $3D$ case and derive the $(3+1)$ dimensional NLSE.
This equation is well known to have solutions representing RWs. As far as I know, there
is no investigation about the IARWs in multi-ion plasma by considering super-thermally distributed electron and positron. Therefore, in this paper, I study the
behavior of the high energetic and giant IARWs in four-component magneto-plasma system.

The manuscript is organized as follows. The magneto-plasma model equations for the nonlinear dynamics
of the IARWs are presented in first part of the Sec. \ref{4sec:Model}. Then the derivation of $(3+1)$-dimensional NLSE, the stability analysis of IAWs and propagation properties of the
IARWs. A conclusions is finally presented in Sec. \ref{4sec:Conclusions}.
\section{Magneto-plasma model, NLSE derivations and Analysis}
\label{4sec:Model}
To analyze the mechanism of MI associated with the fully nonlinear IAWs and the properties of the IARWs we
consider a $3D$ four-component magneto-plasma system containing an opposite polarity two distinct
ion species: positive ions (mass $m_p$; charge $q_p=Z_pe$), negative ions (mass $m_n$; charge $q_n=-Z_ne$),
as well as super-thermal kappa distributed electrons (mass $m_e$; charge $-e$) and positrons (mass $m_{\bar{p}}$; positive
charge $+e$). The charge neutrality condition at equilibrium can be written as
$N_{e0}+Z_nN_{n0}=N_{\bar{p}0}+Z_pN_{p0}$, where $N_{j0}$ denotes the unperturbed (equilibrium) number densities
of species $j$ ($=e,\bar{p},p,$ and $n$ for electrons, positrons, positive ions, and negative ions, respectively);
$Z_p$ ($Z_n$) is the number of protons (electrons) residing on a positive (negative) ions; $e$ is the magnitude
of an electron charge. To account for the importance of the magnetic field, we consider an EUMF
$\textbf{M}=B_0\hat{X}$. Here, $\hat{X}$ is the unit vector along the $X$-axis and
$B_0$ is the strength of the magnetic field.
The three-dimensional model equations of the fully nonlinear IAWs in our electronegative magneto-plasma system is followed by
\begin{eqnarray}
&&\hspace*{-0.5cm} \partial_T N_p+\nabla.(N_p\mathbf{U_p})=0,
\label{4eq:1}\\
&&\hspace*{-0.5cm} \partial_T \mathbf{U_p}+(\mathbf{U_p}.\nabla)\mathbf{U_p}=\omega_{ci}(\mathbf{U_p}\times\hat{X})-\nabla\Psi,
\label{4eq:2}\\
&&\hspace*{-0.5cm} \partial_T N_n+\nabla.(N_n\mathbf{U_n})=0,
\label{4eq:3}\\
&&\hspace*{-0.5cm} \partial_T \mathbf{U_n}+(\mathbf{U_n}.\nabla)\mathbf{U_n}=-\chi\omega_{ci}(\mathbf{U_n}\times\hat{X})+\chi\nabla\Psi,
\label{4eq:4}\\
&&\hspace*{-0.5cm} \nabla^2\Psi=\mu_1N_e-(\mu_1+\mu_2-1)N_{\bar{p}}+\sigma_2N_n-N_p,
\label{4eq:5}
\end{eqnarray}
where $\chi=Z_nm_p/Z_pm_n$, $\mu_1=N_{e0}/Z_pN_{p0}$, and $\mu_2=Z_nN_{n0}/Z_pN_{p0}$. The above mention
dependent/independent variables are normalized as follows: $N_p$, $N_n$, $N_e$, and $N_{\bar{p}}$ are normalized by
$N_{p0}$, $N_{n0}$, $N_{e0}$, and $N_{\bar{p}0}$, respectively.
$\mathbf{U_{p,n}}$ is the velocity of ion fluid which have three components $W_{p,n}$, $U_{p,n}$, and $V_{p,n}$
corresponding to the three space co-ordinates $X$, $Y$, and $Z$. Each velocity components is normalized by the positive ion speed $C_p=(Z_pk_BT_{\bar{p}}/m_p)^{1/2}$; $T_{\bar{p}}$ is the positron temperature; $k_B$ is the Boltzmann constant; $\nabla= (\partial/\partial X,\partial/\partial Y,\partial/\partial Z)$ is a 3D space operator; $T$ is time variable normalized by the ion plasma frequency $\omega_{pp}^{-1}=(m_p/4\pi e^2Z_p^2N_{p0})^{1/2}$; spatial coordinates ($X$, $Y$, $Z$) are normalized by the Debye screening radius $\lambda_{Dp}=(k_BT_{\bar{p}}/4\pi e^2Z_pN_{p0})^{1/2}$; $\Psi$ is electrostatic wave potential normalized by $k_BT_{\bar{p}}/e$; $\omega_{ci}=q_pB_0/m_p$ is the ion cyclotron frequency normalized by $\omega_{pp}$.

The normalized electrons and positrons number densities, which are following the super-thermal kappa distribution, can be written as \cite{Vasyliunas1968,Hellberg2002,Sabry2012,Haque2019,Haque2019a}:
\begin{eqnarray}
&&\hspace*{0.0cm} N_e=\bigg(1-\frac{\mu_3}{\kappa-3/2}\Psi\bigg)^{-\kappa+1/2},
\label{4eq:6}\\
&&\hspace*{0.0cm} N_{\bar{p}}=\bigg(1+\frac{1}{\kappa-3/2}\Psi\bigg)^{-\kappa+1/2},
\label{4eq:7}
\end{eqnarray}
where $\mu_3=T_{\bar{p}}/T_e$, $T_e$ is the temperature of electron, $\kappa$ stand for super thermality parameter.
If $\kappa\rightarrow\infty$), then the super-thermal
distribution function reduces to the well known isothermal (Maxwellian)-distribution and the range of $\kappa$ is $1.6\leq \kappa<6$.
To avoid the complexity  we substitute Eqs. \eqref{4eq:6} and \eqref{4eq:7} into Eq. \eqref{4eq:5}, and expand the
resulting equation up to third order as:
\begin{eqnarray}
&&\hspace*{-0.1cm} \nabla^2\Psi+N_p-\mu_2N_n-1+\mu_2=H_1\Psi+H_2\Psi^2+H_3\Psi^3\cdot\cdot\cdot,
\nonumber\\
\label{4eq:8}
\end{eqnarray}
where
\begin{eqnarray}
&&\hspace*{0.0cm} H_1=\frac{(2\kappa-1)(\mu_1\mu_3+\mu_1+\mu_2-1)}{(2\kappa-3)},
\nonumber\\
&&\hspace*{0.0cm} H_2=\frac{(2\kappa-1)(2\kappa+1)(\mu_1\mu_3^2-\mu_1-\mu_2+1)}{2(2\kappa-3)^2},
\nonumber\\
&&\hspace*{0.0cm} H_3=\frac{(2\kappa-1)(2\kappa+1)(2\kappa+3)(\mu_1\mu_3^3+\mu_1+\mu_2-1)}{6(2\kappa-3)^3}.
\nonumber\
\end{eqnarray}

The D-region is the innermost and F-region is the outermost regions of the Earth's ionosphere.
Many heavy and lighter ion species are present in these regions as $(H^+,O_2^-)$ electronegative
plasma has been found in D-region and $(H^+,H^-)$ plasma has been found in F-region \cite{Sabry2009}. In laboratory magneto-plasma
experiment $(Ar^+,F^-)$ plasma \cite{Nakamura1984} is used.
All examples below are for the $(H^+,~H^-)$, $(H^+,~O_2^-)$, and $(Ar^+,~F^-)$ electronegative plasma systems, where the parameters are:
$Z_p=Z_n=1$, $m_p=m_n=1.00784$u, and $\chi=1.0$ for $(H^+,~H^-)$; $Z_p=1$, $Z_n=16$, $m_p=1.00784$u, $m_n=15.999$u, and $\chi=0.96$ for $(H^+,~O_2^-)$; $Z_p=18$, $Z_n=9$, $m_p=39.948$u, $m_n=18.9984$u, and $\gamma=1.05$ for $(Ar^+,~F^-)$; $\mu_1 = 0.1$-$0.9$, $\mu_2 = 0.1$-$0.9$, $\mu_3 = 0.1$-$0.9$, and $\kappa = 1.6$-$5$. For these set of magneto-plasma parameters, the neutrality condition should be verified.
\\

To construct a nonlinear theory for the propagation of IAWs and IARWs in my considered plasma
systems, I use the reductive perturbation technique \cite{Taniuti1969,Asano1969} (RPT) to derive
a $(3+1)$ dimensional NLSE. Hence, I first introduce the independent variables like in \cite{Haque2019}
$\xi=\varepsilon X$, $\eta=\varepsilon Y$, $\zeta=\varepsilon(Z-v_gT)$, and $\tau=\varepsilon^2T$;
where $v_g$ is the group velocity of IAWs and $\varepsilon~(1>\varepsilon>0)$ is
a small parameter. The dependent variables are approximated via
\begin{eqnarray}
&&\hspace*{-0.3cm} N_{pn}=1+\sum\limits_{m=1}^\infty\varepsilon^{m}\sum\limits_{l=-\infty}^\infty N^{(m)}_{pn l}(\xi,\eta,\zeta,
                        \tau)~e^{i\Gamma l},
\label{4eq:9}
\end{eqnarray}
\begin{eqnarray}
\begin{pmatrix}
               U_{pn} \\
              \\
               V_{pn}
\end{pmatrix}=\sum_{m=1}^\infty \epsilon^{m+1} \sum_{l=-\infty}^\infty
\begin{pmatrix}
              U^{(m)}_{pn l}(\xi,\eta,\zeta,\tau) \\
              \\
              V^{(m)}_{pn l}(\xi,\eta,\zeta,\tau)
\end{pmatrix}~e^{i\Gamma l},
\label{3eq:10}
\end{eqnarray}
\begin{eqnarray}
\begin{pmatrix}
               W_{pn} \\
              \\
               \Psi
\end{pmatrix}=\sum_{m=1}^\infty \epsilon^{m} \sum_{l=-\infty}^\infty
\begin{pmatrix}
              W^{(m)}_{pn l}(\xi,\eta,\zeta,\tau) \\
              \\
              \Psi^{(m)}_l(\xi,\eta,\zeta,\tau)
\end{pmatrix}~e^{i\Gamma l},
\label{4eq:11}
\end{eqnarray}
where, $\Gamma=kX-\omega T$ and $k$($\omega$) denotes the
carrier wavenumber (ion angular frequency). Since $N_{pn}$, $U_{pn}$, $V_{pn}$, $W_{pn}$, and $\Psi$ are real,
the coefficients in Eqs. \eqref{4eq:9}-\eqref{4eq:11} satisfy the condition $\Delta^{(m)}_{-l}=\Delta_l^{(m)^\ast}$,
where $\Delta=(\Psi,N_{pn}, U_{pn}, V_{pn}, W_{pn})$ and the asterisk indicates the complex conjugation.
By substituting the new stretched coordinates and Eqs. \eqref{4eq:9}-\eqref{4eq:11} into Eqs. \eqref{4eq:1}-\eqref{4eq:4} and
\eqref{4eq:8} we collect the coefficients for different powers of $\varepsilon$. I get the first
order quantities for $m=1,~l=1$ as:
\begin{eqnarray}
&&\hspace*{-1.0cm} \Big[N^{(1)}_{p1},~W^{(1)}_{p1}\Big]^T=\Bigg[\frac{k^2}{\omega^2},~\frac{k}{\omega}\Bigg]^T\Psi^{(1)}_1,
\nonumber\\
&&\hspace*{-1.0cm} \Big[N^{(1)}_{n1},~W^{(1)}_{n1}\Big]^T=\Bigg[-\frac{\chi k^2}{\omega^2},~-\frac{\chi k}{\omega}\Bigg]^T\Psi^{(1)}_1,
\nonumber\
\end{eqnarray}
where $T$ denotes the transpose of the matrix. By solving the above equations, I establish a dispersion relation
as follows:
\begin{eqnarray}
&&\hspace*{-1.3cm} \frac{\omega}{k}=\sqrt{\frac{1+\chi\mu_2}{k^2+H_1}}.
\label{4eq:12}
\end{eqnarray}
From Eq. \eqref{4eq:12}, we see that the angular frequency depends on the ion mass ratio $\chi$,
positive and negative ions number density ratio, spectral index $\kappa$, and the
other MPPs. Equation \eqref{4eq:12} is also known as the dispersion relation of the IAWs.
\begin{figure}[h!]
\centering
\includegraphics[width=75mm]{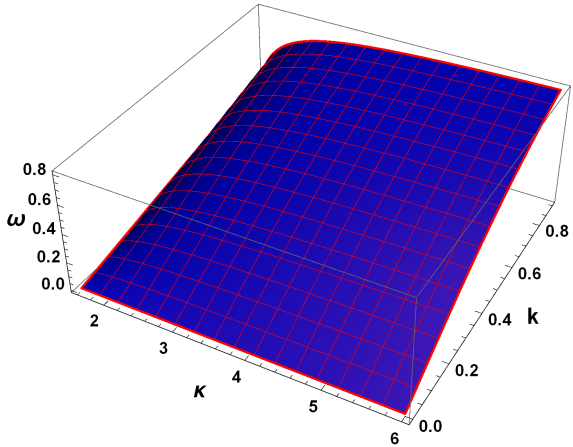}
\caption{For $(H^+,~H^-)$ electronegative plasma system the variation of $\omega$ with $k$ and $\kappa$ when $\chi=1.0$, $\mu_1=0.8$, $\mu_2=0.4$, and
         $\mu_3=0.7$.}
 \label{4Fig:F1}
\end{figure}

I have graphically shown how the angular frequency varies with $\kappa$ and $k$ for $(H^+,~H^-)$ plasma
system in Fig. \ref{4Fig:F1}. It is seen the angular frequency increases with $k$ significantly, but it is increasing for the low values $\kappa$ then become constant. It is evident from Eq. \eqref{4eq:12} that the ion cyclotron frequency
has no effect on the dispersive curve.

Again for $(m=2)$, $(l=1)$ I get the following relations:
\begin{eqnarray}
&&\hspace*{-0.1cm} i\omega N_{p1}^{(2)}=ik W_{p1}^{(2)}-v_g\partial_{\zeta} N_{p1}^{(1)}+\partial_{\zeta} W_{p1}^{(1)},
\nonumber\\
&&\hspace*{-0.1cm} i\omega W_{p1}^{(2)}=ik \Psi_{1}^{(2)}+\partial_{\zeta} \Psi_{1}^{(1)}-v_g\partial_{\zeta} W_{p1}^{(1)},
\nonumber\\
&&\hspace*{-0.1cm} i\omega N_{n1}^{(2)}=ik W_{n1}^{(2)}-v_g\partial_{\zeta} N_{n1}^{(1)}+\partial_{\zeta} W_{n1}^{(1)},
\nonumber\\
&&\hspace*{-0.1cm} i\omega W_{n1}^{(2)}=-ik\chi \Psi_{1}^{(2)}-\chi\partial_{\zeta} \Psi_{1}^{(1)}-v_g\partial_{\zeta} W_{n1}^{(1)},
\nonumber\\
&&\hspace*{-0.1cm} N_{p1}^{(2)}+2i k\partial_{\zeta}\Psi_1^{(1)}= k^2\Psi_1^{(2)}+H_1\Psi_1^{(2)}+\mu_2N_{n1}^{(2)}.
\nonumber\
\end{eqnarray}
Using these above equations, I obtain the group velocity of IAWs as:
\begin{eqnarray}
&&\hspace*{-1.3cm} v_g\big(=\partial_k\omega\big)=\frac{\omega\big(1+\mu_2\chi-\omega^2\big)}{k(1+\chi\mu_2)}.
\label{4eq:13}
\end{eqnarray}

\begin{figure}[h!]
\centering
\includegraphics[width=75mm]{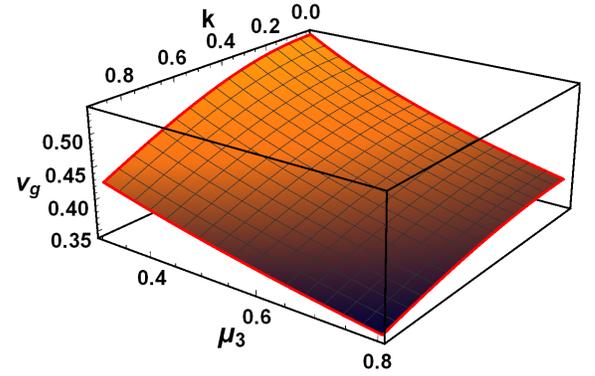}
\caption{For $(H^+,~H^-)$ electronegative plasma system the variation of $v_g$ with k and $\mu_3$ when $\chi=1.0$, $\mu_1=0.8$, $\mu_2=0.4$, and
         $\kappa=1.6$.}
 \label{4Fig:F2}
\end{figure}
The mathematical expression of the group velocity of the IAWs is represented in eq. \eqref{4eq:13}
and it is also evident that the ion cyclotron
frequency has no effect on the group velocity ($v_g$). But $v_g$ of the IAWs depends on the positive/negative
ion mass ratio $\chi$, wave number $k$, negative ions to positive ions densities ratio $\mu_2$, and other plasma parameters.
Figure \ref{4Fig:F2} shows that $v_g$ is negatively affected by the temperature ratio $\mu_3$ and
$k$ in $(H^+,~H^-)$ plasma medium. It means that the group velocity linearly decreases with $\mu_3$ and $k$.
It is also true for both ($H^+,~O_2^-$) and ($Ar^+,~F^-$) plasma media.

Now, for $m=2$, $l=2$ I get the reduced equations
\begin{eqnarray}
\left.
      \begin{aligned}
           &&\hspace*{0.0cm} N^{(2)}_{p2}=H_4^{(22)}|\Psi^{(1)}_1|^2,~~~~W^{(2)}_{p2}=H_5^{(22)}|\Psi^{(1)}_1|^2\\
           &&\hspace*{0.0cm} N^{(2)}_{n2}=H_6^{(22)}|\Psi^{(1)}_1|^2,~~~~W^{(2)}_{n2}=H_7^{(22)}|\Psi^{(1)}_1|^2\\
           &&\hspace*{-6.0cm} \Psi^{(2)}_2=H_8^{(22)}|\Psi^{(1)}_1|^2
      \end{aligned}
\right\}
\qquad
\label{4eq:14}
\end{eqnarray}
where,
\begin{eqnarray}
&&\hspace*{-0.3cm} H_4^{(22)}=\Big[2k^2 H_8^{(22)}\omega^2+3k^4\Big]\times\Big[2\omega^4\Big]^{-1},
\nonumber\\
&&\hspace*{-0.3cm} H_5^{(22)}=\Big[2\omega^2 k H_8^{(22)}+k^3\Big]\times\Big[2\omega^3\Big]^{-1},
\nonumber\\
&&\hspace*{-0.3cm} H_6^{(22)}=\Big[k^2\chi\big(3k^2\chi-2 H_8^{(22)}\omega^2\big)\Big]\times\Big[2\omega^4\Big]^{-1},
\nonumber\\
&&\hspace*{-0.3cm} H_7^{(22)}=\Big[k^3\chi^2-2k\chi H_8^{(22)}\omega^2\Big]\times\Big[2\omega^3\Big]^{-1},
\nonumber\\
&&\hspace*{-0.3cm} H_8^{(22)}=-\Big[2\omega^4H_2+3k^4\big(\chi^2\mu_2-1\big)\Big]\times\Big[6k^2\omega^4\Big]^{-1}.
\nonumber\
\end{eqnarray}
Furthermore, the second-order $(m=2)$ quantities in the zeroth harmonic $(l=0)$, arising from the nonlinear self-interaction
of the carrier waves, can be expressed as
\begin{eqnarray}
\left.
      \begin{aligned}
           &&\hspace*{0.0cm} N^{(2)}_{p0}=H_9^{(22)}|\Psi^{(1)}_1|^2,~~~~W^{(2)}_{p0}=H_{10}^{(22)}|\Psi^{(1)}_1|^2\\
           &&\hspace*{0.0cm} N^{(2)}_{n0}=H_{11}^{(22)}|\Psi^{(1)}_1|^2,~~~~W^{(2)}_{n0}=H_{12}^{(22)}|\Psi^{(1)}_1|^2\\
           &&\hspace*{-6.0cm} \Psi^{(2)}_0=H_{13}^{(22)}|\Psi^{(1)}_1|^2
      \end{aligned}
\right\}
\qquad
\label{4eq:15}
\end{eqnarray}
where
\begin{eqnarray}
&&\hspace*{-0.3cm} H_9^{(20)}=\Big[H_{13}^{(20)}\omega^3+2v_gk^3+k^2\omega\Big]\times\Big[v_g^2\omega^3\Big]^{-1},
\nonumber\\
&&\hspace*{-0.3cm} H_{10}^{(20)}=\Big[H_{13}^{(20)}\omega^2+k^2\Big]\times\Big[v_g\omega^2\Big]^{-1},
\nonumber\\
&&\hspace*{-0.3cm} H_{11}^{(20)}=\Big[2v_gk^3\chi^2+k^2\chi^2\omega-\chi H_{13}^{(20)}\omega^3\Big]
\times\Big[v_g^2\omega^3\Big]^{-1},
\nonumber\\
&&\hspace*{-0.3cm} H_{12}^{(20)}=\Big[k^2\chi^2-\chi H_{13}^{(20)}\omega^2\Big]\times\Big[v_g\omega^2\Big]^{-1},
\nonumber\\
&&\hspace*{-0.3cm} H_{13}^{(20)}=\Big[2v_g^2H_2\omega^3+2v_gk^3\big(\chi^2\mu_2-1\big)+k^2\omega\big(\chi^2\mu_2-1\big)\Big]
\nonumber\\
&&\hspace*{3.3cm} \times\Big[\omega^3\big(1+\chi\mu_2-v_g^2H_1\big)\Big]^{-1}.
\nonumber\
\end{eqnarray}
Finally, by taking the third order $(m=3)$ with the first harmonic mode $(l=1)$ and using
Eqs. \eqref{4eq:12} -\eqref{4eq:15}, I obtain a nonlinear partial differential equation in
the form of the (3+1)-dimensional NLSE \cite{Guo2014}:
\begin{eqnarray}
&&\hspace*{-1.0cm} i\partial_{\tau}\Phi+P\partial^2_{\zeta\zeta}\Phi+Q|\Phi|^2\Phi-R\bigg(\partial^2_{\xi\xi}\Phi+\partial^2_{\eta\eta}\Phi\bigg)=0,
\label{4eq:16}
\end{eqnarray}
where $\Phi=\Psi^{(1)}_1$ for simplicity. The term $P$ ($R$) is known as longitudinal (transverse) dispersive coefficient and $Q$ is
known as non-linear term. The coefficients $P$, $R$, and $Q$ can be defined as:
\begin{eqnarray}
&&\hspace*{-0.0cm} P\Big(=\frac{1}{2}\partial_k v_g\Big)=\frac{3v_g}{2k\omega}\big(\omega\chi\mu_2-\omega^3-v_gk\big),
\nonumber\\
&&\hspace*{0.0cm}R=\frac{\omega^3}{2k^2(\chi\mu_2+1)}\Bigg[\frac{\omega^2-\chi^2\omega_{ci}^2+\chi\mu_2\big(\omega^2-\omega_{ci}^2\big)}
{\big(\omega_{ci}^2-\omega^2\big)\big(\omega^2-\chi^2\omega_{ci}^2\big)}+1\Bigg],
\nonumber\\
&&\hspace*{-0.0cm} Q=\bigg(3\omega^3H_3+2\omega^3H_2\Big(H_8^{(22)}+H_{13}^{(20)}\Big)\bigg)\times\Big(2k^2(1+
\nonumber\\
&&\hspace*{0.0cm} \chi\mu_2)\Big)^{-1}-k\Big(H_5^{(22)}+\chi\mu_2H_7^{(22)}+H_{10}^{(20)}+\chi\mu_2H_{12}^{(20)}\Big)
\nonumber\\
&&\hspace*{0.7cm} \times\big(1+\chi\mu_2\big)^{-1}-\omega\Big(H_4^{(22)}+\chi\mu_2H_6^{(22)}+H_9^{(20)}
\nonumber\\
&&\hspace*{2.7cm} +\chi\mu_2H_{11}^{(20)}\Big)\times\big(2+2\chi\mu_2\big)^{-1}.
\nonumber\
\end{eqnarray}
Recent studies confirm that the transverse perturbation with external uniform magnetic field is the leading cause to introduce the
transverse dispersion term $R$ \cite{Guo2014,Haque2019,Haque2019a,Haque2020}. It is obvious that
the coefficients $P$, $R$, and $Q$ are the function of relevant plasma parameters, such as, positive to negative ion mass
ratio $\chi$, ratio of the products of negative ion charge and number density to positive ion charge and number
density $\mu_2$, ion cyclotron frequency $\omega_{ci}$, wave number $k$, and so on.
\\

\begin{figure}[h!]
\centering
  \subfloat[]{\includegraphics[height=75mm, width=80mm]{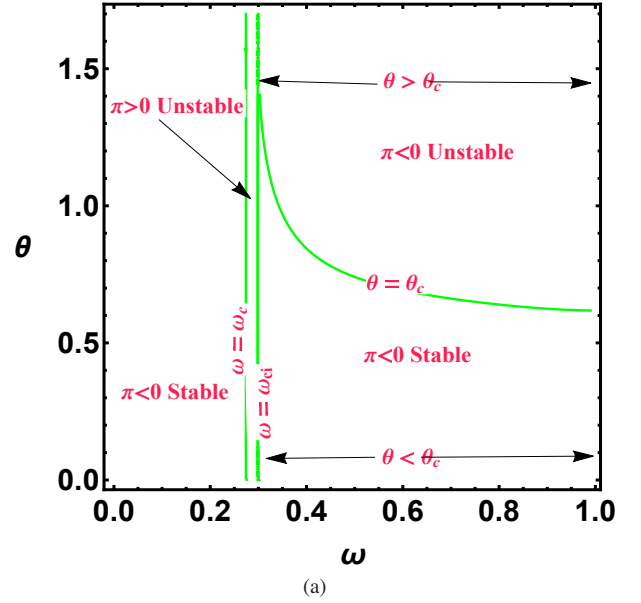}}
   \hfill
  \subfloat[]{\includegraphics[height=75mm, width=80mm]{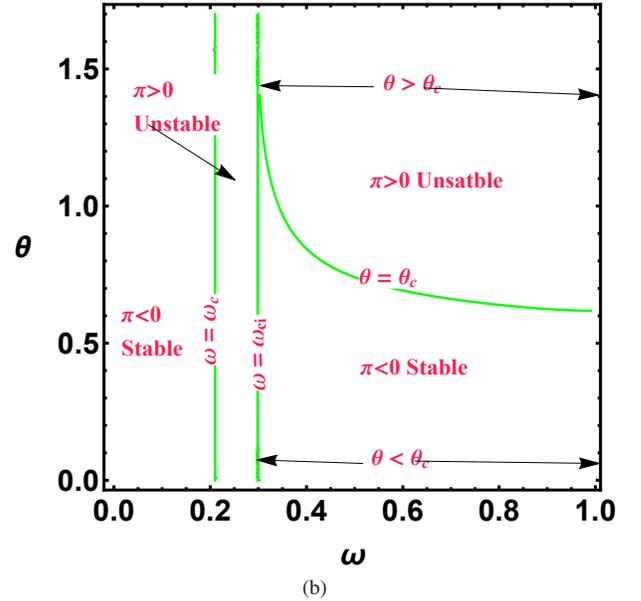}}
\caption{The frequency regimes of stability ($\pi<0$) and instability ($\pi>0$) in $(H^+,~H^-)$ electronegative plasma
         for different values of $\kappa$: (a) $\kappa=1.6$ and
   (b) $\kappa=7$ when the other parameters are $\chi=1.0$, $\mu_1=0.8$, $\mu_2=0.4$, $\omega_{ci}=0.3$, and $\mu_3=0.6$.}
 \label{4Fig:F3}
\end{figure}

\begin{figure*}[t!]
\centering
\subfloat[]{\includegraphics[width=53mm, height=33mm]{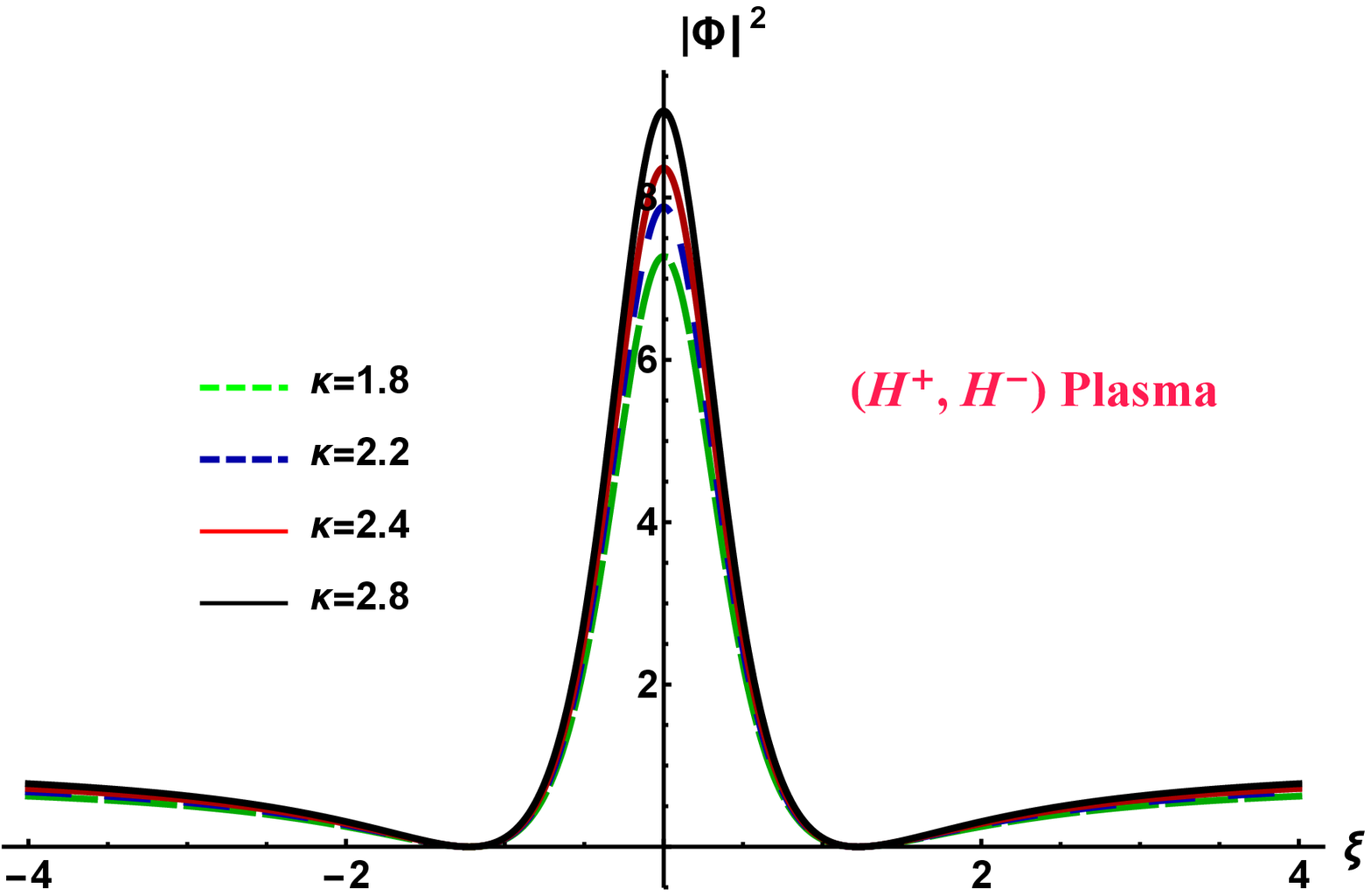}}
\hfill
\subfloat[]{\includegraphics[width=53mm, height=33mm]{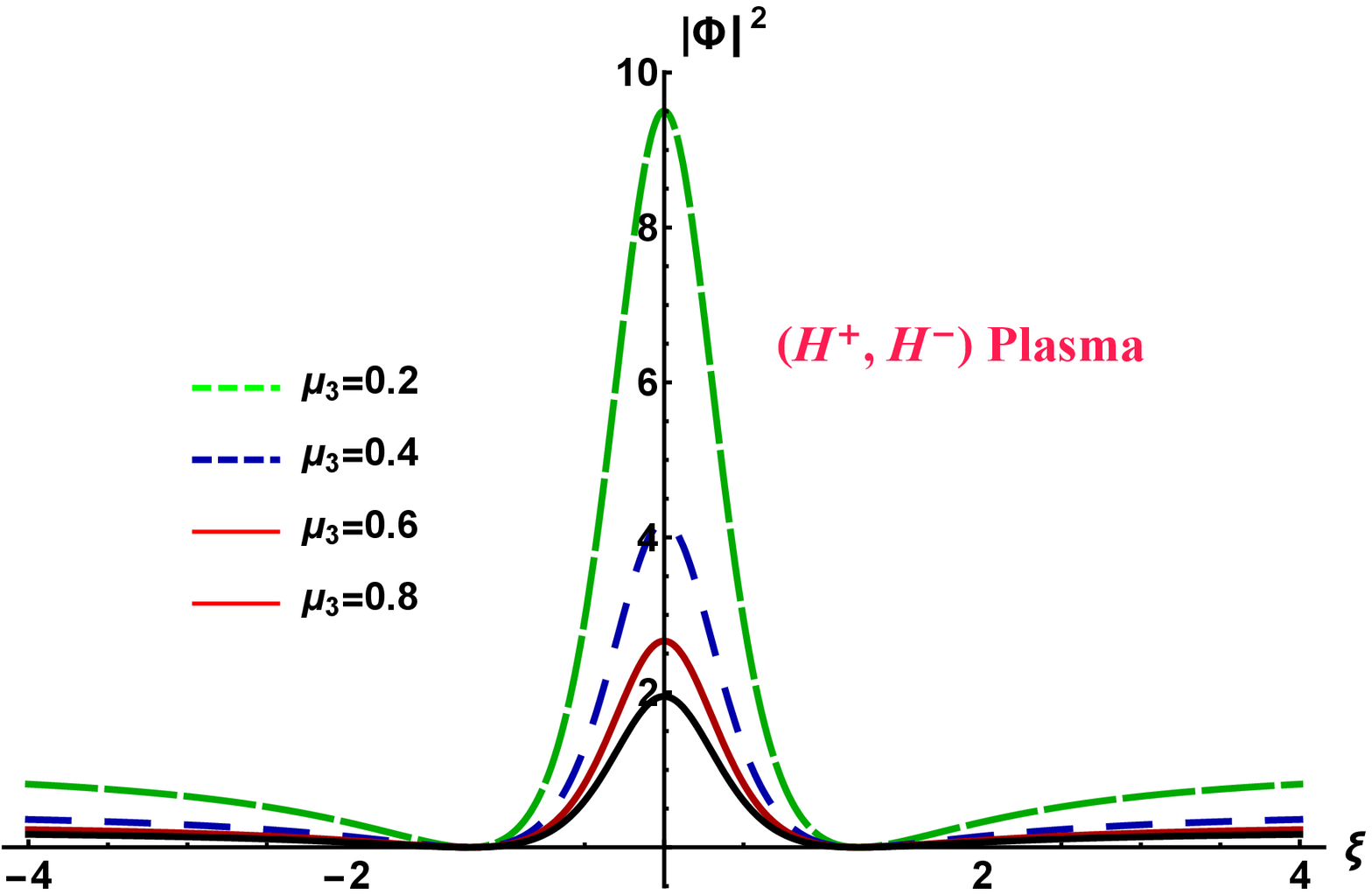}}
\hfill
\subfloat[]{\includegraphics[width=53mm, height=33mm]{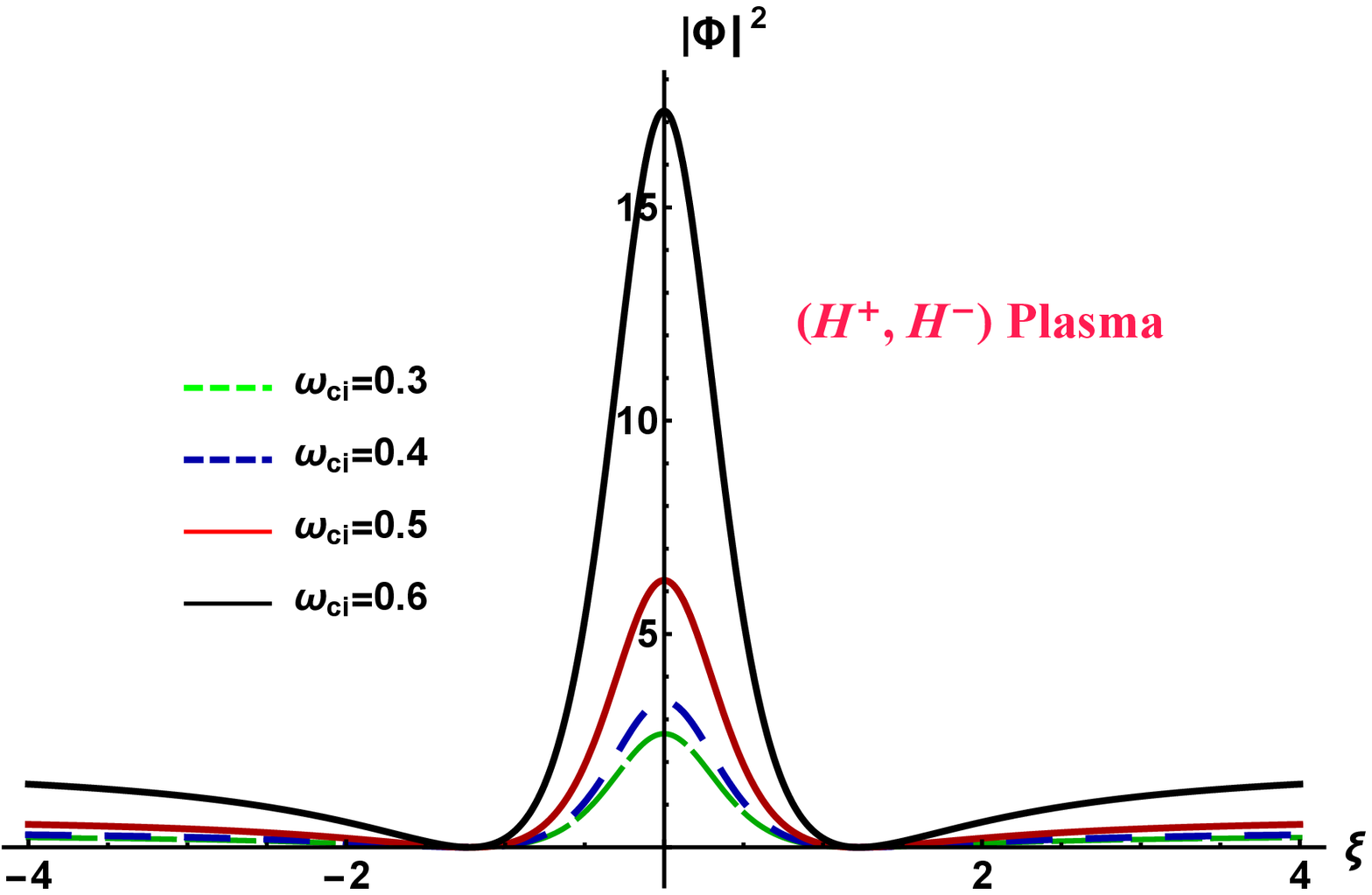}}
\caption{The variation of the first-order IARWs [represented by Eq. \eqref{4eq:18}] formed in the $(H^+,~H^-)$ electronegative plasma system is shown against $\xi$ at different (a) $\kappa$ with $\mu_3=0.6$,
$k=0.65$, and $\omega_{ci}=0.3$; (b) $\mu_3$ with $\kappa=2$, $k = 0.8$, and $\omega_{ci}=0.3$; (c) $\omega_{ci}$ with $\kappa=2$, $k = 0.8$, and $\mu_3=0.6$. Here, the other parameters are $\mu_1=0.8$, $\mu_2=0.4$, $\eta=0$, $\zeta=0$, and $\tau=0$.}
\label{4Fig:F5}
\end{figure*}
Now, I will analyze the MI of IAWs, which is described by the 3D NLSE Eq. \eqref{4eq:16},
and the properties of first- and second-order IARWs.
By considering a plane wave solution $\Phi=\Phi_0\exp(-i Q|\Phi_0|^2\tau)$ of Eq. \eqref{4eq:16}, one can
obtain the nonlinear dispersion relation for modulated IAWs as \cite{Shalini2017,Haque2020}:
\begin{eqnarray}
&&\hspace*{-1.3cm} \Omega^2=K^4\Big(\frac{\sigma_2/P}{1+M_{\theta}^2}\Big)^2\times\bigg[1-\frac{2|\Phi_0|^2
\big(1+M_\theta^2\big)\sigma_1}{K^2\pi_2}\bigg].
\label{4eq:17}
\end{eqnarray}
Here, $\sigma_1=PQ$, $\sigma_2= P^2M_\theta^2-PR$, $\Phi_0$ is the perturbed amplitude, and $\Omega$ ($\ll\omega$)
and $K$ \Big($\equiv \sqrt{K_\xi^2+K_\eta^2+K_\zeta^2}$\Big) are the normalized modulated wave frequency
and wave number, respectively. Here, $(K_\xi,~K_\eta,~K_\zeta)$ are components of modulated wave number
$K$ along the stretched coordinates $(\xi,~\eta,~\zeta)$.
The parameter $L_\theta \Big(= K_\xi/\sqrt{K_\eta^2+K_\zeta^2}\Big)$
is connected with the modulational obliqueness $\theta$ ($\theta$ represents the angle between wave
vector $\textbf{K}$ and the resultant $K_\zeta\zeta$ and $K_\eta\eta$) as $\theta=\arctan(M_\theta)$.
Equation \eqref{4eq:17} indicates that a critical wave number ($K_c$) exists when $K^2<K_c^2=2|\Phi_0|^2\big(1+L_\theta^2\big)\Lambda$,
where $\pi=\sigma_1/\sigma_2$, and for this condition $\theta$ has a critical value
through $\theta_c\equiv \arctan\big(\sqrt{R/P}\big)$.
Now, according to \eqref{4eq:17}, the MI may occur for $K_c^2 > K^2$ if one of the following conditions is satisfied:
$\sigma_1<0$ and $\sigma_2<0$ or $\sigma_1>0$ and $\sigma_2>0$.
Therefore, $\sigma_1$ and $\sigma_2$ are mainly responsible for the MI. The unstable region ($\pi > 0$) exists when $\sigma_1$ and
$\sigma_2$ are same sign and the region becomes stable ($\pi<0$) when $\sigma_1$ and $\sigma_2$ are opposite signs.
Further, for the condition $K_c^2 > K^2$ the decay/growth rate can be gained from \eqref{4eq:17} as
$\Upsilon=Im~\Omega=(K^2\pi_2/P+PM_{\theta}^2)\times\sqrt{K^{-2}K_c^2-1}$.
If the condition $PK_\xi^2-M(K_\eta^2+K_\zeta^2)=Q~|\phi_0|^2$ is satisfied, then the maximum decay/growth rate is obtained
as $\Upsilon_{max}=Q~|\phi_0|^2$.

Figures 3 shows the stable and unstable frequency regimes
in $\omega - \theta$ plane for different values of the spectral index $\kappa$ at $\omega_{ci} = 0.3$.
Figure \ref{4Fig:F3} depicts the stable and unstable regions for the $(H^+,~H^-)$ electronegative plasma
system (i.e., $\chi = 1$) under consideration. I have not reported the MI analyses in $(H^+,O_2^-)$ and $(Ar^+,F^-)$
electronegative plasma systems since they are found to be almost identical to those shown in Fig. \ref{4Fig:F3}.
There are several stability and instability frequency regimes, which are separated by the lines of $\omega = \omega_c$,
$\omega= \omega_{ci}$, and $\theta=\theta_c$ in $\omega-\theta$ plane as illustrated in Fig \ref{4Fig:F3}.
It is clear from the Fig. \ref{4Fig:F3}a that at $\omega < \omega_c$ modulated IAW is stable
since the critical parameter $\pi$ (i.e., $\sigma_1 <0$ and $\sigma_2 > 0$) is negative, but modulated IAWs becomes
unstable for $\omega_c<\omega<\omega_{ci}$ since the MI criterion $\pi >0$ (i.e., $\sigma_1 > 0$ and $\sigma_2 > 0$) is satisfied.
It means that the MI is independent of the obliqueness $\theta$ in the region $\omega_c<\omega<\omega_{ci}$. Note that for $\omega < \omega_{ci}$
the MI criterion is similar to that in unmagnetized $1$D case. On the other hand, for $\omega>\omega_{ci}$ the MI is strongly dependent of the
obliqueness $\theta$. Here, $\theta = \theta_c$ line divides the region $\omega>\omega_{ci}$ into stable and unstable regions. For $\omega>\omega_{ci}$,
the critical parameter $\pi$ (i.e., $\sigma_1 >0$ and $\sigma_2 > 0$) is positive and the MI occurs for $\theta>\theta_c$. On the contrary,
modulated IAWs are stable for $\theta < \theta_c$ because MI parameter $\pi$ (i.e., $\sigma_1 >0$ and $\sigma_2 < 0$) is negative.
It should be recalled that the critical frequency $\omega_c$ switches to the lower value as $\kappa$ increases, as depicted in Fig. \ref{4Fig:F3}b.
Finally, the parameter $\omega_c$ plays an important role for the super-thermal parameter $\kappa$.

\begin{figure}[h!]
\centering
  \subfloat[]{\includegraphics[height=75mm, width=80mm]{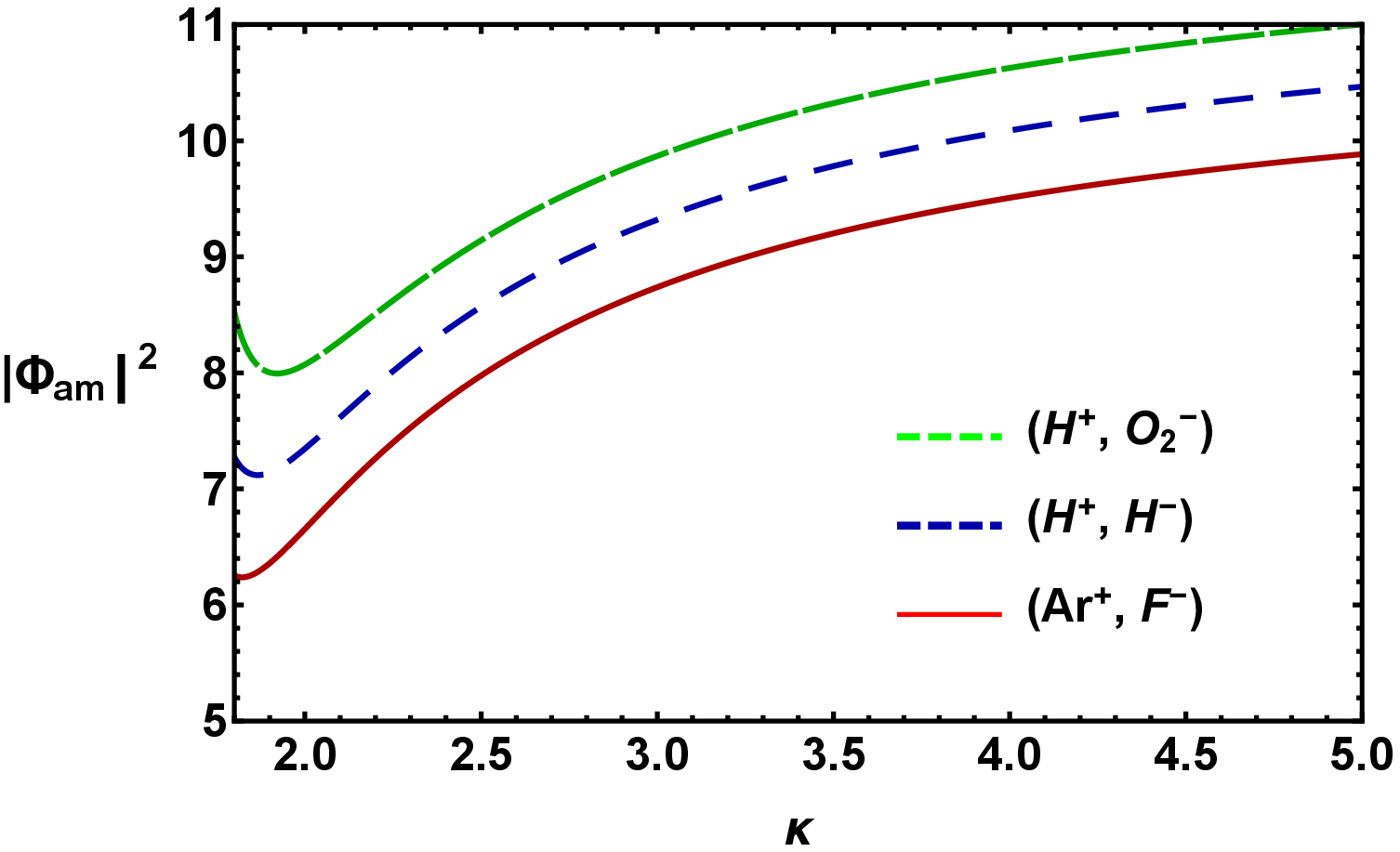}}
   \hfill
  \subfloat[]{\includegraphics[height=75mm, width=80mm]{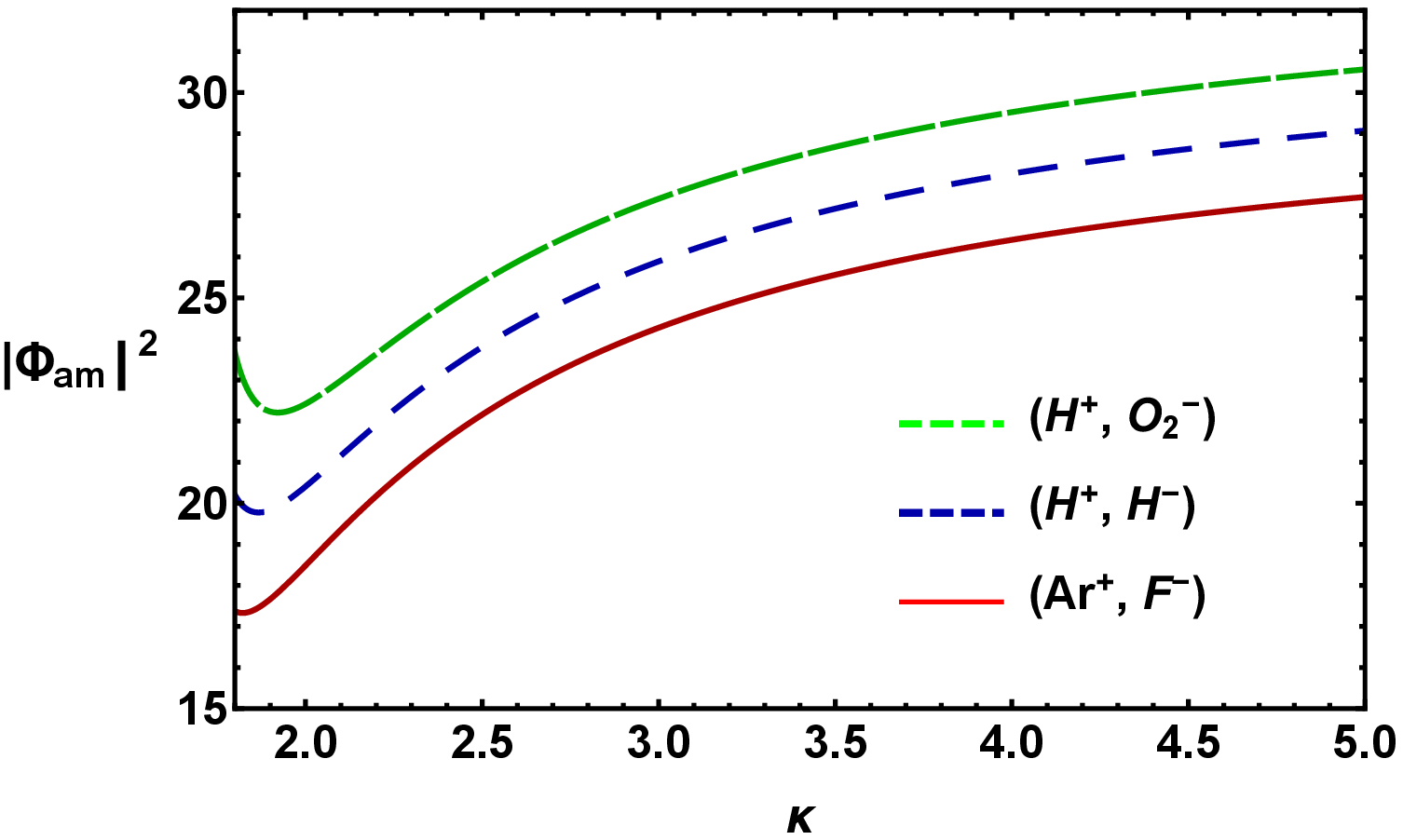}}
\caption{The first-(a) [represented by Eq. \eqref{4eq:18}] and second-(b) [represented by Eq. \eqref{4eq:19}] order IARWs amplitude variation against the spectral index $\kappa$ in the $(H^+,O_2^-)$, $(H^+,~H^-)$, and $(Ar^+,F^-)$ electronegative plasma
systems at $\zeta=0$, $\eta=0$, and $\xi=0$ with $\mu_1=0.8$, $\mu_2=0.4$, $\mu_3=0.6$, $k=0.65$, and $\omega_{ci}=0.3$.}
\label{4Fig:F6}
\end{figure}

Now, I will study the propagation properties of the first- and second-order IARWs in different types of
electronegative plasmas in the presence of external uniform magnetic field $B_0$.
The first- and second-order IARW solutions of the $(3+1)$-dimensional NLSE \eqref{4eq:16}
can be presented, respectively, as \cite{Guo2014}:
\begin{equation}\label{4eq:18} \Phi_1=\sqrt{\lambda Q^{-1}}~e^{i\tau\lambda}\times(\chi-8i\tau\lambda-3)\times(1+\Re_1)^{-1},
\end{equation}
\begin{equation}\label{4eq:19}\
\Phi_2=\sqrt{\lambda Q^{-1}}~(\Re_4+\Re_2+\Re_3)\times \Re_4^{-1}~e^{i\tau\lambda},
\end{equation}
where $\Re_1=2\delta^2+4\tau^2\lambda^2$, $\lambda=2Q+P$, $\delta=\xi+\eta+\zeta$. Other parameters
$\Re_2$, $\Re_3$, and $\Re_4$ are
\begin{eqnarray}
&&\hspace*{-0.5cm} \Re_2=-\big(18\tau^2\lambda^2+3\delta^2+20\tau^4\lambda^4\big)
\nonumber\\
&&\hspace*{+0.5cm} \times~2^{-1}-\big(\delta^4+12\delta^2\tau^2\lambda^2\big)\times~2^{-1}+3/2,
\nonumber\
\end{eqnarray}
\begin{eqnarray}
&&\hspace*{-0.5cm} \Re_3=\big(2\tau^2\lambda^2-3\delta^2+4\tau^4\lambda^4+4\delta^2\tau^2\lambda^2+\delta^4-15/4\big)\tau\lambda,
\nonumber\
\end{eqnarray}
\begin{eqnarray}
&&\hspace*{-0.1cm} \Re_4=3/32+\frac{33\Theta}{4}+\delta^4\Theta+18\Theta^2-\bigg(3\delta^2\Theta-\frac{\delta^4}{8}
\nonumber\\
&&\hspace*{0.1cm} -\frac{16\Theta^3}{3}-\frac{\delta^6}{12}-4\delta^2\Theta^2-\frac{9\delta^2}{16}\bigg),
\nonumber\
\end{eqnarray}
where $\Theta=\tau^2\lambda^2/2$. Figure \ref{4Fig:F5} reveals the variation of the nonlinear first-order IARWs formed in the $(H^+,H^-)$ electronegative plasma media in the presence of the transverse perturbation. I now examine the effect of the super-thermal parameter $\kappa$, the positron to electron temperature ratio $\mu_3$, and the positive ion cyclotron frequency $\omega_{ci}$ on the properties of first-order IARWs in the $(H^+,H^-)$ electronegative plasma system. It is seen that the profiles of IARW solution of \eqref{4eq:18} are significantly modified by the above mentioned parameters. It shows that as I increase $\kappa$, $\mu_3$, and $\omega_{ci}$, the amplitude and width of the first-order IARWs increase with $\kappa$, decrease with $\mu_3$, and increase with $\omega_{ci}$,respectively. The physics of this result is
that since the IARWs are assumed to be associated with the nonlinearity of our plasma system, the amplitude and width of the IARWs increase. The physical parameters (i.e., $\kappa$ and $\omega_{ci}$) enhance the nonlinearity of the $(H^+,H^-)$ electronegative plasma media but $\mu_3$ suppress the nonlinearity. It is worth to mention that the second-order IARWs have the similar qualitative behaviour of first-order IARWs at the same parameters. Note also that the same qualitative results are also found in the case of $(H^+,O_2^-)$ and $(Ar^+,F^-)$ electronegative plasma systems. Due to this reason, we do not include these results for simplicity.

The amplitude variation of the first and second-order IARW in the $(H^+,O_2^-)$, $(H^+,H^-)$, and $(Ar^+,F^-)$ electronegative
plasma systems are graphically presented in Fig. 5. The amplitude preciously depends on the nonlinearity of the plasma system
so that the amplitude of IARWs increases (decreases) with increasing (decreasing) the nonlinearity. Figure 6a and 6b show that the amplitude decreases with decreasing the value of $\kappa$. However, the nonlinearity of the systems increases with decrease in the value of the positive to negative ion mass
ratio $\chi$ that means at the same value of the plasma parameters the nonlinearity in the $(H^+,H^-)$ electronegative plasma
is higher than that of $(Ar^+,F^-)$ electronegative plasma, but is smaller than that of $(H^+,O_2^-)$ electronegative plasma.

\begin{figure}[h!]
\centering
  \subfloat[]{\includegraphics[height=75mm, width=80mm]{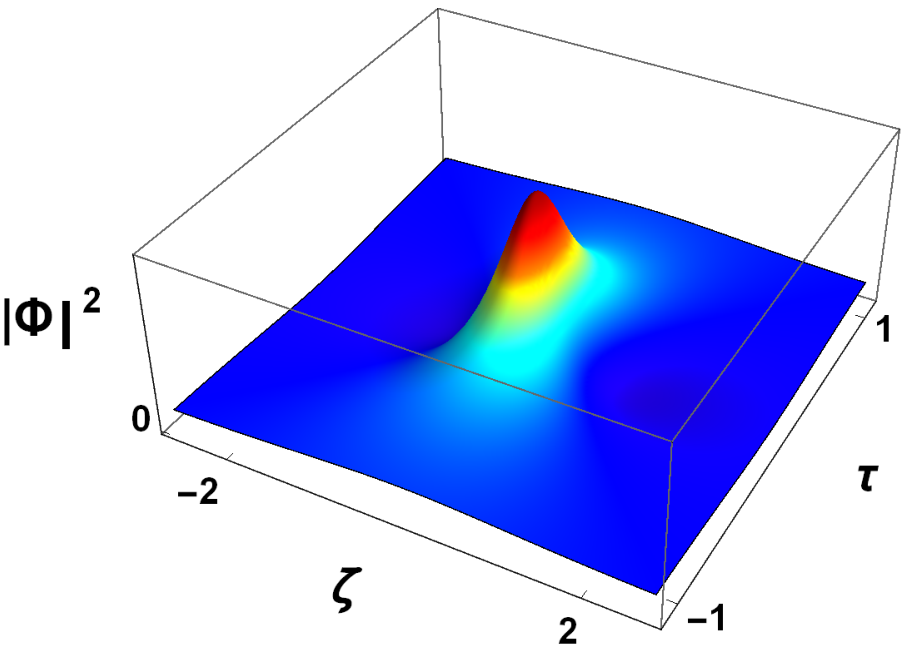}}
   \hfill
  \subfloat[]{\includegraphics[height=75mm, width=80mm]{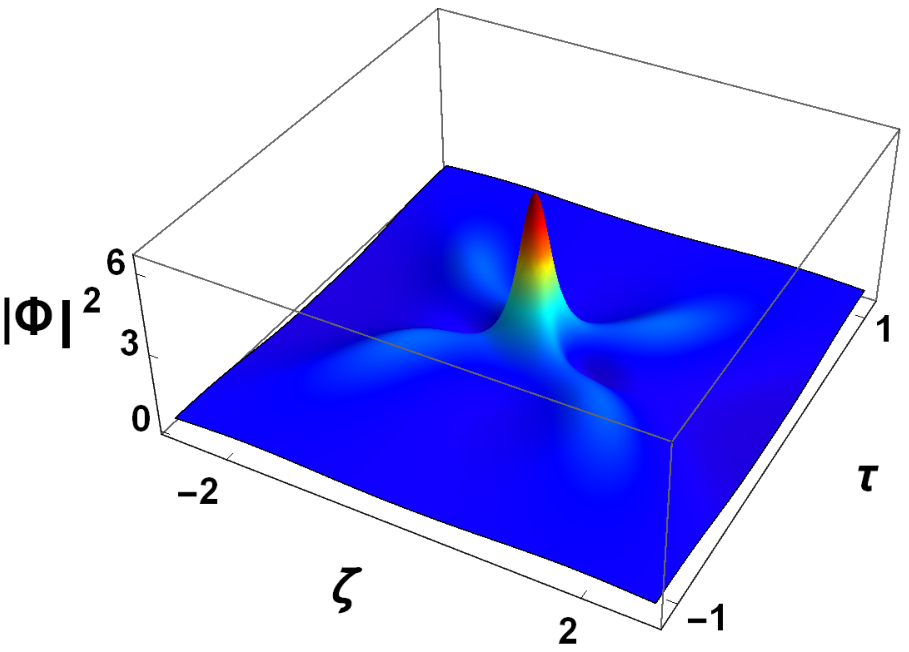}}
\caption{The first-(a) [represented by Eq. \eqref{4eq:18}] and second-(b) [represented by Eq. \eqref{4eq:19}] order IARWs formed in the $(H^+,~H^-)$ electronegative plasma system are shown against $\zeta$ and $\tau$ at $\eta=0$ and $\xi=0$ with $\mu_1=0.8$, $\mu_2=0.4$, $\mu_3=0.6$, $\kappa=2$, $k=0.8$, and $\omega_{ci}=0.3$.}
\label{4Fig:F7}
\end{figure}

Figure \ref{4Fig:F7} shows the spatiotemporal evolution of the first- and second-order IARWs formed in the $(H^+,H^-)$ electronegative plasma system at $\eta = 0$ and $\xi = 0$. As I see from these figures that our plasma system can concentrate a large amount of IAW wave energy in a relatively small region. Note that the amplitude of the second-order IARWs (as shown in Fig. \ref{4Fig:F7}b) is almost three times of the first-order IARWs (as shown in Fig. \ref{4Fig:F7}a). It means that the second-order IARWs can absorb more energy from the background waves and concentrate more energy into relative small region rather than the first-order IARWs. Note also that the second-order IARWs are more spiky (i.e., sharper amplitude and narrower width) than the first-order IARWs. It is clearly seen that second-order IARWs involve much more complicated nonlinear profiles. Of course, such solutions could not be seen if I neglect the transverse perturbations $R$, i.e., for 1D NLSE.

\section{Conclusions}
\label{4sec:Conclusions}
 I have theoretically investigated the three dimensional modulation of IAWs and the properties of the high energetic IARWs in a magnetized plasma system containing the electrons and positrons following super-thermal kappa distribution function and the opposite polarity ions. A $(3+1)$-dimensional NLSE, which is well known to have solutions representing the RWs in the unstable region, has been derived by using the reductive perturbation method. It is found that due to the transverse plane perturbations, the modulation properties of the IAWs in unmagnetized and magnetized plasma systems are different. It is noted that the strong magnetic field under consideration significantly modifies the modulational instability/stability regions and the basic features (viz., amplitude, width, and speed) of IARWs. The effect of the different physical parameters (viz., super-thermal index $\kappa$, the positron to electron temperature ratio $\mu_3$, the positive ion cyclotron frequency $\omega_{ci}$, etc.) on the modulation of IAWs and IARWs are reported.

 It is expected that my present work is useful in understanding the physics of nonlinear phenomena like modulational instability of ion-acoustic waves and rogue waves in the D-F regions of Earth's ionosphere [e.g., ($H^+,~H^-$), ($H^+,~O_2^-$) electronegative plasma] \cite{Coates2007,Massey1976}
and laboratory device [e.g., $(Ar^+,F^-)$ electronegative plasma] \cite{Nakamura1984}.
\section*{Data Availability Statement}
\textcolor{blue}{Data sharing is not applicable to this article as no new data were created or analyzed in this study.}


\begin{thebibliography}{99}
\bibitem{Vladimirov2003} S. V. Vladimirov, K. Ostrikov, M. Y. Yu, and G. E. Morfill, Phys. Rev. E \textbf{2003}, 67, 036406.

\bibitem{Misra2008} A. P. Misra and P. K. Shukla, Phys. Plasmas \textbf{2008}, 15, 122107.

\bibitem{Tie-Lu2015} L. Tie-Lu, W. Yun-Liang, and L. Yan-Zhen, Chin. Phys. B \textbf{2015}, 24, 025202.

\bibitem{Panguetna2018} C. B. Tabi, C. S. Panguetna, T. C. Kofan\'{e}, Physica B: Condensed Matter \textbf{2018}, 545, 370.

\bibitem{Sabry2009} R. Sabry, W. M. Moslem, and P. K. Shukla, Phys. Plasmas \textbf{2009}, 16, 032302.

\bibitem{Abdelwahed2016} H. G. Abdelwahed, E. K. El-Shewy, M. A. Zahran, and S. A. Elwakil, Phys. Plasmas \textbf{2016}, 23, 022102.

\bibitem{Selim2016} M. M. Selim, Eur. Phys. J. Plus \textbf{2016}, 131, 93.

\bibitem{Chowdhury2017} N. A. Chowdhury, A. Mannan, M. M. Hasan, and A. A. Mamun, Chaos \textbf{2017}, 27, 093105.

\bibitem{Massey1976} H. Massey, \textit{Negative Ions}, 3rd ed. (Cambridge University Press, Cambridge, 1976).

\bibitem{Khondaker2019} S. Khondaker, A. Mannan, N. A. Chowdhury, and A. A. Mamun, Contributions to Plasma Phys. \textbf{2019}, 59, e201800125.

\bibitem{Mishra2002} M. K. Mishra, A. K. Arora, and R. S. Chhabra, Phys. Rev. E \textbf{2002}, 66, 046402.

\bibitem{Mishra1995} M. K. Mishra, R. S. Chhabra, and S. R. Sharma, Phys. Rev. E \textbf{1995}, 51, 4790.

\bibitem{Pedersen1965} A. Pedersen, Tellus \textbf{1965}, 17, 02-45.

\bibitem{Coates2007} A. J. Coates, F. J. Crary, G. R. Lewis, D. T. Young, J. H. Waite Jr., and E. C. Sittler, Geophys. Res. Lett. \textbf{2007}, 34, L22103.

\bibitem{Chaizy1991} P. H. Chaizy, H. Reme, J. A. Sauvaud, C. Duston, R. P. Lin, D. E. Larson, D. L. Mitchell, K. A. Anderson, C. W. Carlson,
                    A. Korth, and D. A. Mendis, Nature (London) \textbf{1991}, 349, 393.

\bibitem{Jacquinot1977} J. Jacquinot, B. D. McVey, and J. E. Scharer, Phys. Rev. Lett. \textbf{1977}, 39, 88.


\bibitem{Hansen1988} E. T. Hansen and A. G. Emslie, The Physics of Solar Flares (Cambridge University Press, Cambridge, 1988).

\bibitem{Adriani2009} O. Adriani, G. C. Barbarino, and G. A. Bazilevskaya, Nature \textbf{2009}, 458, 607.

\bibitem{Surko1989} C. M. Surko, M. Leventhal, and A. Passner, Phys. Rev. Lett. \textbf{1989}, 62, 901.
\bibitem{Greaves1994} R. G. Greaves, M. D. Tinkle, and C. M. Surko, Phys. Plasmas \textbf{1994}, 1, 1439.

\bibitem{Chowdhury2017a} N. A. Chowdhury, A. Mannan, and A. A. Mamun, Phys. Plasmas \textbf{2017}, 24, 113701.

\bibitem{Vasyliunas1968} V. M. Vasyliunas, J. Geophys. Res. \textbf{1968}, 73, 2839.
\bibitem{Hellberg2002} M. A. Hellberg and R. L. Mace, Phys. Plasmas \textbf{2002}, 09, 1495.
\bibitem{Sabry2012} R. Sabry, W. M. Moslem, and P. K. Shukla, Phys. Plasmas \textbf{2012}, 19, 122903.
\bibitem{Asbridge1968} J. R. Asbridge, S. J. Bame, and I. B. Strong, J. Geophys. Res. \textbf{1968}, 73, 5777.
\bibitem{Cairns1995} R. A. Cairns, A. A. Mamun, R. Bingham, R. Bostrom, R. O. Dendy, C. M. C. Nairn, and P. K. Shukla, Geophys. Res. Lett. \textbf{1995}, 22, 2709.
\bibitem{Leubner1982} M. P. Leubner, J. Geophys. Res. \textbf{1982}, 87, 6335.
\bibitem{Armstrong1983} T. P. Armstrong, M. T. Paonessa, E. V. II. Bell, and S. M. Krimigis, J. Geophys. Res. \textbf{1983}, 88, 8893.
\bibitem{Futaana2003} Y. Futaana, S. Barabasha, M. Holmstr\"{o}ma, and A. Bhardwajb, J. Geophys. Res. \textbf{2003}, 108, 151.
\bibitem{Shahmansouri2013} M. Shahmansouri and H. Alinejad, Phys. Plasmas \textbf{2013}, 20, 082130.


\bibitem{Mei2016} S. Guo, L. Mei, Y. He, and Y. Li, Plasma Phys. Control. Fusion \textbf{2016}, 58, 025014.

\bibitem{Guo2012} S. Guo, L. Mei, and A. Sun, Annals of Phys. \textbf{2012}, 332, 38.

\bibitem{Guo2014} S. Guo and L. Mei, Phys. Plasmas \textbf{2014}, 21, 082303.

\bibitem{Haque2019} M. N. Haque, A. Mannan, and A. A. Mamun, Plasma Phys. Rep. \textbf{2019}, 45, 1026.
\bibitem{Haque2019a} M. N. Haque, A. Mannan, and A. A. Mamun, Contributions to Plasma Phys. \textbf{2019}, 59, e201900049.
\bibitem{Haque2020} M. N. Haque and A. Mannan, IEEE Trans. Plasma Sci. \textbf{2020}, 48, 2591.
\bibitem{Haque2021} M. N. Haque and A. Mannan, Contributions to Plasma Phys. \textbf{2021}, 61, e202000161.
\bibitem{Jahan2021} S. Jahan, M. N. Haque, N. A. Chowdhury, A. Mannan and A. A. Mamun, Universe \textbf{2021}, 7, 63.
\bibitem{Nakamura1984} Y. Nakamura and I. Tsukabayashi, Phys. Rev. Lett. \textbf{1984}, 52, 2356.

\bibitem{Taniuti1969} T. Taniuti and N. Yajima, J. Math. Phys. \textbf{1969}, 10, 1369.

\bibitem{Asano1969} N. Asano, T. Taniuti, and N. Yajima, J. Math. Phys. \textbf{1969}, 10, 2020.



\end{thebibliography}
\end{document}